\documentclass[preprint,5p,times,twocolumn]{elsarticle}

\usepackage{amssymb}
\usepackage{units}
\usepackage{siunitx}

\usepackage{lineno}
\usepackage{textcomp} 
\journal{Nuclear Instruments and Methods in Physics Research}

\begin{document}

\begin{frontmatter}



\title{Schlieren Imaging for the Determination of the Radius of an Excited Rubidium Column}
\author[CERN,MPP,TUM]{A.-M. Bachmann}
\ead{anna-maria.bachmann@cern.ch}
\author[MPP]{M. Martyanov}

\author[MPP]{J. Moody}
\author[HHU]{A. Pukhov}
\author[MPP,CERN]{P. Muggli}
\address[CERN]{CERN, Geneva, Switzerland}
\address[MPP]{Max-Planck Institute for Physics, Munich, Germany}
\address[TUM]{Technical University Munich, Munich, Germany}
\address[HHU]{Heinrich-Heine-Universit\"at D\"usseldorf, D\"usseldorf, Germany}

\begin{abstract}
AWAKE develops a new plasma wakefield accelerator using the CERN SPS proton bunch as a driver \cite{AWAKE}. The proton bunch propagates through a $10 \, \textnormal{m}$ long rubidium plasma, induced by an ionizing laser pulse. The co-propagation of the laser pulse with the proton bunch seeds the self modulation instability of the proton bunch that transforms the bunch to a train with hundreds of bunchlets which drive the wakefields. Therefore the plasma radius must exceed the proton bunch radius. Schlieren imaging is proposed to determine the plasma radius on both ends of the vapor source. We use Schlieren imaging to estimate the radius of a column of excited rubidium atoms. A tunable, narrow bandwidth laser is split into a beam for the excitation of the rubidium vapor and for the visualization using Schlieren imaging. With a laser wavelength very close to the D2 transition line of rubidium ($\lambda \approx 780 \, \textnormal{nm}$), it is possible to excite a column of rubidium atoms in a small vapor source, to record a Schlieren signal of the excitation column and to estimate its radius. We describe the method and show the results of the measurement. 
\end{abstract}


\end{frontmatter}
\section{Introduction}
\label{sec:intro}
\begin{figure}[htb!]
\centering
	 	\includegraphics[width =0.8\columnwidth,trim={0cm 25.9cm 8cm 0cm},clip]{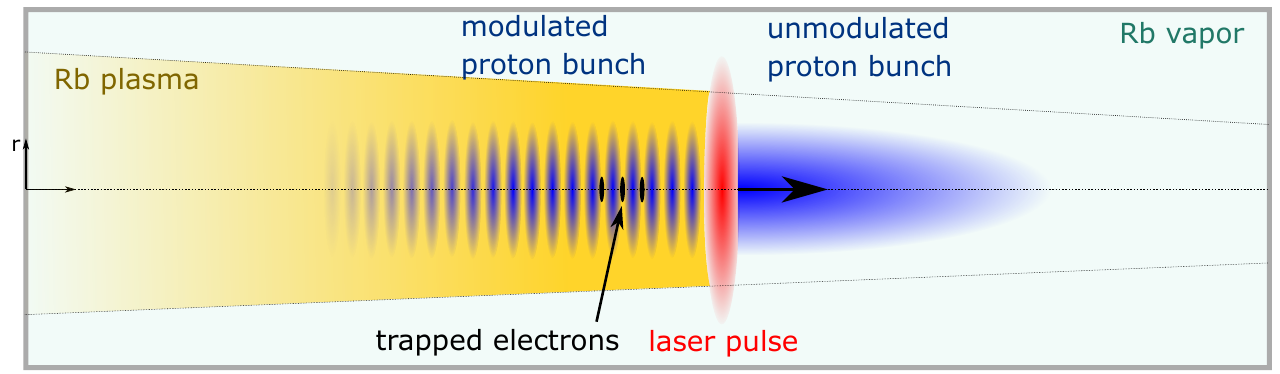}
		\caption{Sketch of the AWAKE principle}
\label{fig:sketchofawakeprinciple}
\end{figure}

In AWAKE the CERN SPS proton bunch propagates through a $10 \, \textnormal{m}$ long rubidium (Rb) vapor source. An ionizing laser is co-propagating with the proton bunch, so that the front half of the bunch travels through Rb vapor and the back half of the bunch interacts with the Rb plasma. The creation of the hard edge plasma seeds the self-modulation of the particle bunch in a plasma, which divides the $12\, \textnormal{cm}$ long proton bunch into micro bunches at the plasma period ($\lambda_{pe} \approx 1 \, \textnormal{mm}$). The bunchlets modulate the electron plasma density and drive the wakefield resonantly. It is foreseen to accelerate injected electrons in the wakefield \cite{AWAKE}. The principle of the AWAKE project is sketched in figure \ref{fig:sketchofawakeprinciple}.\\
Since the proton bunch has to propagate inside the plasma channel a diagnostic is necessary to show that the plasma radius is larger than the proton bunch radius ($\sigma_{r} \approx \SI{200}{\micro\metre}$). We propose Schlieren imaging to determine the plasma radius. In this paper we report on using Schlieren imaging to measure the radius of a column of excited Rb atoms \cite{bachmannMasterthesis}. The experiment serves as a preparation for the plasma radius measurement. 
\section{Method}
\label{sec:method} 
\begin{figure}[htb!]
\centering
	 	\includegraphics[width =0.6\columnwidth,trim={0cm 20.6cm 1.5cm 0cm},clip]{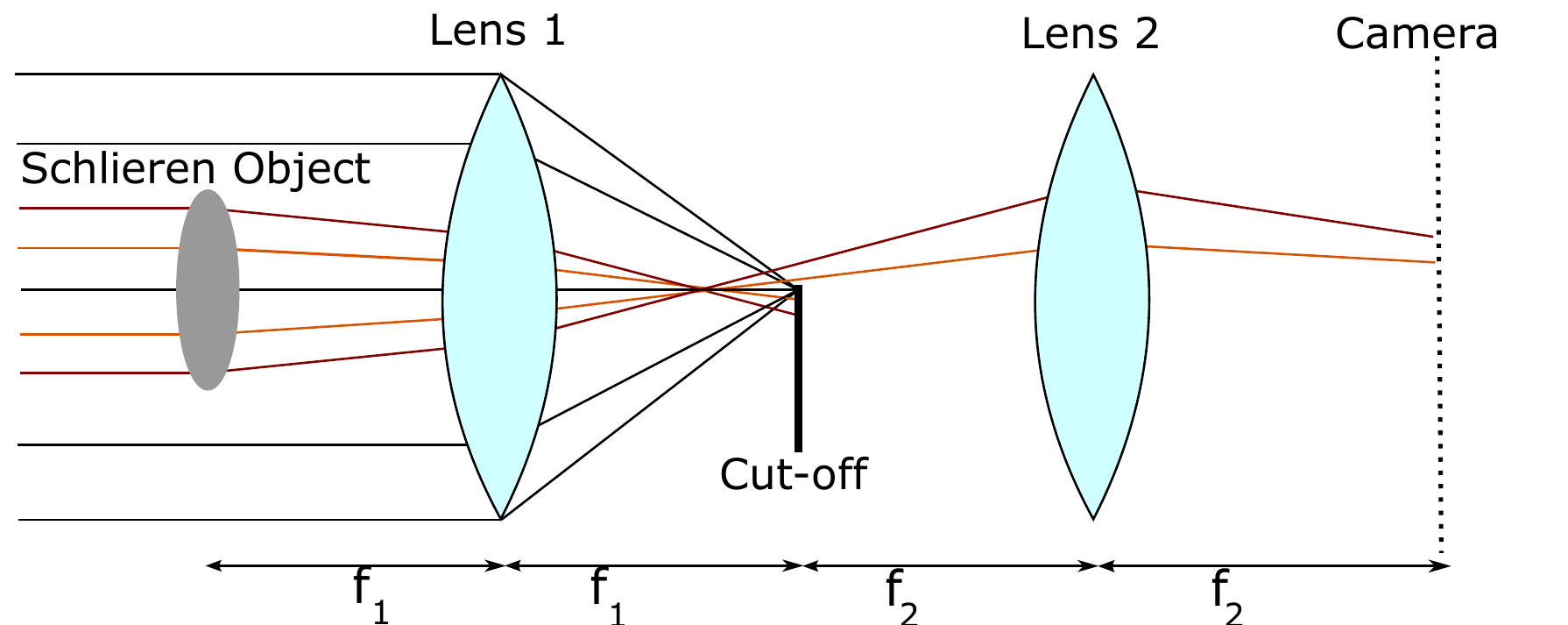}
		\caption{Sketch of the Schlieren imaging principle with $f_1$ and $f_2$ the focal lengths of lens 1 and lens 2 \cite{schlierenSpringer}}
\label{fig:sketchofschlierensetup}
\end{figure}

The principle of Schlieren imaging is sketched in figure \ref{fig:sketchofschlierensetup}. The setup consists of two focusing lenses and an opaque object (cut-off) blocking light at the focal point of the first lens. A beam of collimated light propagates through the setup covering the transparent object to be visualized and that is placed in front of the first lens. The two lenses are arranged to form an image of the object onto the camera. The rays propagating through the object (orange and red in figure \ref{fig:sketchofschlierensetup}) are slightly bent by the transparent medium with a different refractive index than the surrounding medium. While the unbent rays (black) cross each other at the first lens focal point, the bent rays cross each other at a shifted position along the optical axis. Thus unbent rays can be blocked with a cut-off, while rays bent away from the cut-off can pass. The shift of the bent rays depends on the difference in the refractive index between the imaged object and the surrounding medium and on the propagation distance inside the object.\\
Here we use this method to image a column of excited Rb atoms surrounded by atoms in the ground state. The refractive index of vapor for light with a frequency $\omega_{L}$ near the medium optical transition frequency $\omega_{ij}$ from the lower state $i$ to the upper state $j$ is \cite{laserspektroskopie}
\begin{equation}
n_{vapor} =\nobreak \sqrt[]{1+ \frac{N_i \, e^2}{\epsilon_0 \, m_e} \, \sum_{j\neq i} \frac{f_{ij}}{\omega_{ij}^2 - \omega_L^2 + i \gamma_{ij} \omega_L} }, 
\label{eq:refractiveindexvapor}
\end{equation}
with $N_i$ the density of atoms in the lower state $i$, $f_{ij}$ the oscillator strength of the transition and $\gamma_{ij} =  1/ \tau_{ij}$ with $\tau_{ij}$ the lifetime of the upper state $j$, $e$ the elementary charge, $\epsilon_0$ the vacuum permittivity and $m_e$ the electron mass. 
Due to the resonant denominator, the index of refraction can be very different from one when $\omega_L^2 \simeq \omega_{ij}^2$, i.e. when the laser frequency is close to the transition frequency (anomalous dispersion). 
At room temperature the vast majority of the electrons are in the lower state, contributing to an index of refraction very different from one. Rubidium has a single electron on its external O shell, therefore $N_i$ is equal to the Rb vapor density. Excitation of the transition with a near resonant frequency laser beam depopulates the ground state and thus changes the index of refraction of the pumped volume. At low pumping intensity the change in index is proportional to the laser beam intensity. At higher intensity saturation occurs with the maximum depopulation reaching at most $50 \%$ of the original population, when using a continuous wave (CW) laser.\\
\begin{figure}[htb!]
\centering
		\includegraphics[width =0.6\columnwidth,trim={0 15cm 0 0cm},clip]{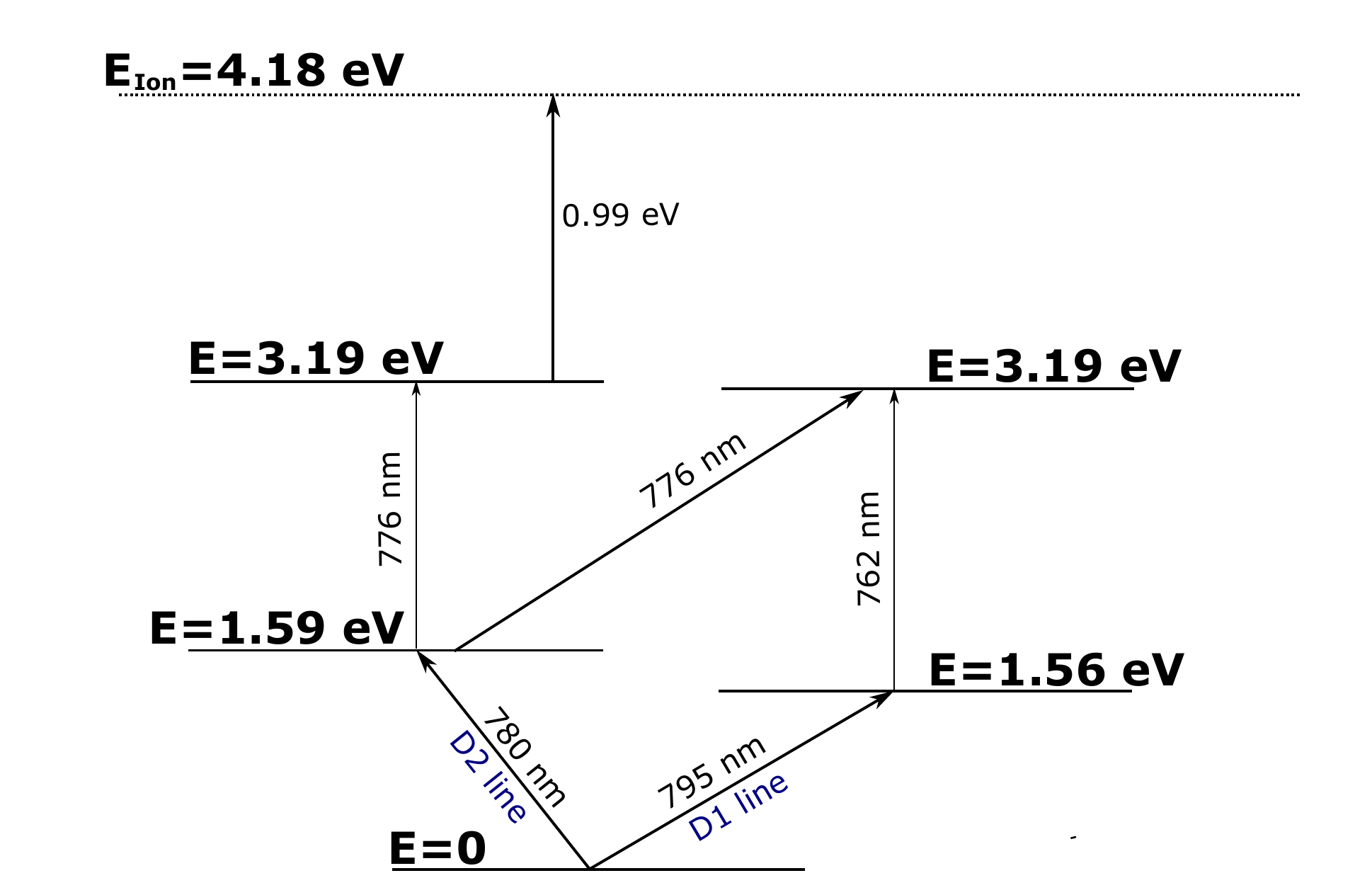}
		\caption{Simplified atomic level structure of Rb with transitions near the wavelength of interest (D2 line) \cite{rbdata2}}
		\label{fig:rbenergylevels}
\end{figure}
Figure \ref{fig:rbenergylevels} shows a simplified atomic level structure of Rb. Rubidium has a transition line at $\lambda \approx 780 \, \textnormal{nm}$ from the ground to the first excited state (D2 line). This transition is easily accessible with commercial lasers. For the radius measurement a CW laser (DLC DL pro from TOPTICA) is used. It is tunable from $763.1 \, \textnormal{nm}$ to $ 813.6 \, \textnormal{nm}$, has a line width of less than $1 \, \textnormal{MHz}$ and a maximum power of $138 \, \textnormal{mW}$. The laser can be fine tuned over a range of $50 \, \textnormal{GHz}$ by applying voltage onto a piezo element that changes the laser cavity length.\\
The laser frequency is determined with absorption spectroscopy before the radius measurement of the excited column.
\begin{figure}[htb!]
\centering
		\includegraphics[width =1\columnwidth,trim={0 4.2cm 1cm 0cm},clip]{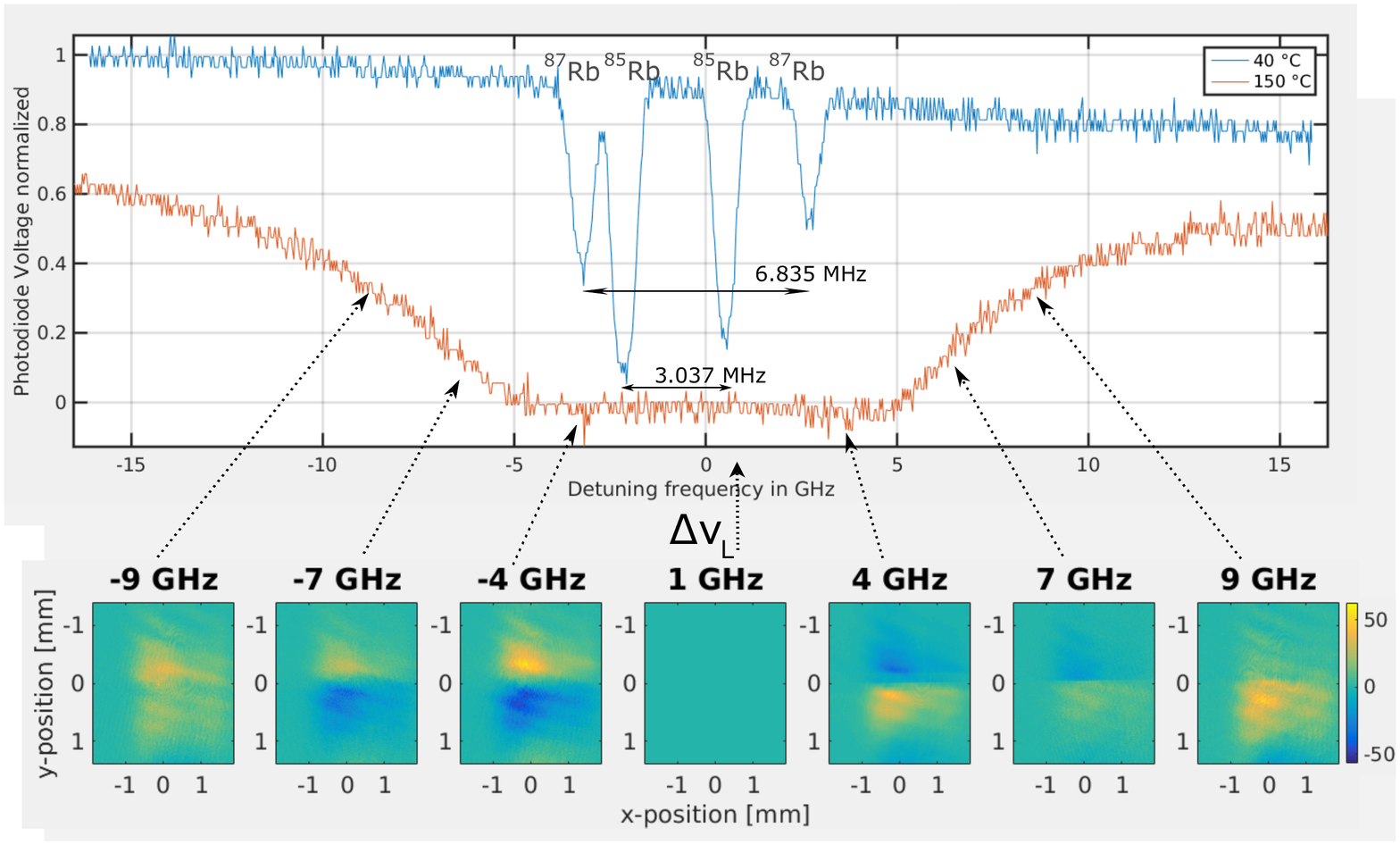}
		\caption{Absorption spectrum near the Rb D2 line for two different Rb cube temperatures (top) and Schlieren signal of the excited Rb column with a horizontal razor blade as cut-off at high density ($T = 150 ^{\circ}C$) for different laser detuning frequencies $\Delta \nu_L$ (bottom)}
		\label{fig:schlierenfrequencydependent}
\end{figure}
The top of figure \ref{fig:schlierenfrequencydependent} shows the absorption spectrum near the Rb D2 line obtained by scanning the piezo voltage and recording with a photodiode the laser intensity transmitted through the Rb vapor. Natural Rb is composed of two isotopes, $^{85}Rb$ with an abundance of $72 \%$ and $^{87}Rb$ with an abundance of $28 \%$. Each of the ground states is split into two hyperfine states. The four lines arising in the absorption spectrum can be used for the laser frequency determination on a very precise level as the frequency distances between the hyperfine states are very well known. They are $6.835 \, \textnormal{GHz}$ for $^{87}Rb$ and $3.037 \, \textnormal{GHz}$ for $^{85}Rb$ \cite{rbdata}.\\
\begin{figure}[htb!]
\centering
		\includegraphics[width =0.8\columnwidth,trim={0cm 17.8cm 3cm 0},clip]{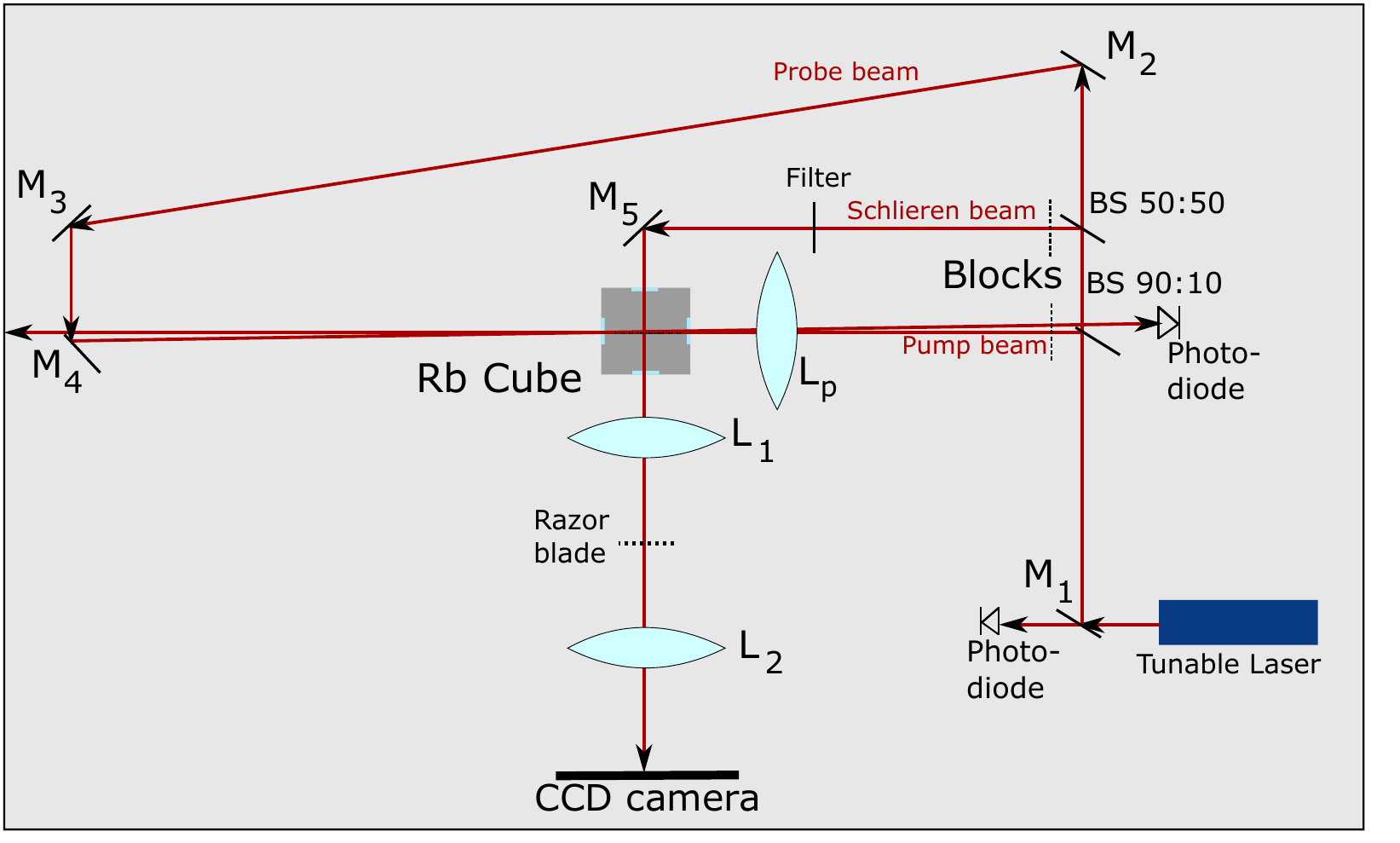}
		\caption{Sketch of the setup for Schlieren imaging of an excited Rb column with absorption spectroscopy for the laser frequency determination}
\label{fig:sketchofsetupschlierenatawake}
\end{figure}
Figure \ref{fig:sketchofsetupschlierenatawake} shows the experimental setup. The vacuum cube containing the Rb can be heated with electrical tapes to produce Rb vapor. The vapor density can be calculated from the vapor temperature and the vapor pressure expression \cite{rbdensitytemperature}. For instance a vapor temperature of $T = 150\, ^{\circ}C$ corresponds to a Rb vapor density of $n_{Rb} \approx 1 \cdot 10^{14} \textnormal{cm}^{-3}$. To avoid saturation of the absorption lines and to resolve the hyperfine structure of the atoms ground states, the absorption spectrum for the laser frequency determination must be recorded at a lower Rb density ($T = 40 ^{\circ}C$, blue line in figure \ref{fig:schlierenfrequencydependent}) before further heating up the source ($T = 150\, ^{\circ}C$, red line in figure \ref{fig:schlierenfrequencydependent}) to reach a sufficiently high Rb vapor density for a detectable Schlieren signal.\\ 
As shown in the setup of figure \ref{fig:sketchofsetupschlierenatawake} the tunable laser is used for excitation (pump beam) and imaging (Schlieren beam): $90 \%$ of the beam intensity is focused with a lens (focal length $f_{L_p}= 200 \, \textnormal{mm}$) onto the center of the Rb cube as pump beam. With a 50:50 beam splitter the remaining $10 \%$ is divided into the Schlieren beam and a probe beam for the laser frequency determination. The Schlieren beam also excites the medium, but due to its much lower intensity (additional filtering before propagation through the Rb vapor) than that of the pump beam, the index of refraction the Schlieren beam experiences is dominated by the effect of the pump beam. For the laser frequency determination the probe beam propagates through the Rb vapor anti parallel to the pump beam and onto a photodiode for absorption spectroscopy. The signal is normalized with a second photodiode placed at the exit of the laser to take into account intensity fluctuations while changing the laser frequency. The Schlieren beam propagates transversely to the pump beam through the vapor and into the Schlieren setup (see figure \ref{fig:sketchofschlierensetup}). The Schlieren and the pump beam can be blocked independently in order to obtain four different images with the CCD camera: both beams blocked, only the Schlieren beam blocked, only the pump beam blocked or none of the beams blocked. The probe beam for absorption spectroscopy is blocked during the Schlieren measurement, since the piezo voltage - laser frequency dependency was determined at low Rb vapor density.\\
\section{Results}
\begin{figure}[htb!]
\centering
		\includegraphics[width =0.8\columnwidth,trim={3cm 17.8cm 6.7cm 0cm},clip]{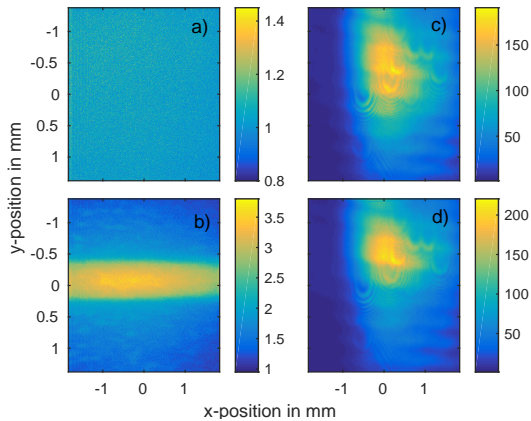}
		\caption{CCD camera images in the four blocking configurations at high density ($T = 150 ^{\circ}C$) and a laser detuning frequency $ \Delta \nu_L = - 8.7 \, \textnormal{GHz}$: a) Schlieren and pump beam blocked, b) only Schlieren beam blocked, c) only pump beam blocked, d) neither Schlieren nor pump beam blocked}
		\label{fig:schlierenshadowgraphycomparison4conf_knifeedge}
\end{figure}

Figure \ref{fig:schlierenshadowgraphycomparison4conf_knifeedge} shows images obtained with the four beam configurations. 
Fifteen images are averaged in each case. During the acquisition of the various images no slow laser intensity variation was observed. A horizontal razor blade (cut-off) blocks half of the Schlieren beam from below the focal point of the first lens, see figure \ref{fig:sketchofschlierensetup}. Figure \ref{fig:schlierenshadowgraphycomparison4conf_knifeedge}a was recorded blocking both beams, i.e. this image corresponds to the camera background and as expected shows no features. When recording the image in figure \ref{fig:schlierenshadowgraphycomparison4conf_knifeedge}b the Schlieren beam was blocked. Thus this image shows the fluorescence of the column of excited Rb atoms. It shows the pump beam propagating horizontally, with the highest intensity reached at the beam waist ($x \simeq 0$) as expected. Note the low intensity of this signal with respect to that of figure \ref{fig:schlierenshadowgraphycomparison4conf_knifeedge}c and \ref{fig:schlierenshadowgraphycomparison4conf_knifeedge}d. The Schlieren beam is recorded head-on covering the waist of the pump beam when present. Figure \ref{fig:schlierenshadowgraphycomparison4conf_knifeedge}c shows the image of the Schlieren beam when the pump beam is blocked, i.e. without the presence of a transparent object (the excited Rb vapor column). This image shows the light of the Schlieren beam that is not blocked by the razor blade, which is slightly less than half of the unblocked Schlieren beam. Figure \ref{fig:schlierenshadowgraphycomparison4conf_knifeedge}d is the image with both beams propagating through the Rb vapor, i.e. the Schlieren image.\\
In order to extract the image of the region that was excited by the pump beam, image a is subtracted from each of the other three images. Then images b and c are subtracted from image d. The resulting image is called the ''Schlieren signal'' in the following.\\
Figure \ref{fig:schlierenfrequencydependent} shows the Schlieren signal for different laser detuning frequencies $\Delta \nu_L = (\omega_L - \omega_{ij})/2\pi $, $ \Delta \nu_L =0$ corresponds to the center of the four D2 transition lines. With a horizontally oriented razor blade as cut-off, the undeflected and the downwards deflected rays are blocked. This razor blade geometry displays a vertical index gradient in the object and is suitable for imaging the effect of the cylindrical, horizontally propagating laser beam of this experiment. For a negative (positive) detuning frequency $\Delta \nu_L$, the Schlieren beam is defocused (focused) by the excited Rb column, whose refractive index is smaller (larger) than one, as the surrounding medium (see equation \ref{eq:refractiveindexvapor}). Due to the blocking of downwards deflected rays by the razor blade this leads to a decrease (increase) of light on the camera in the lower half of the column and an increase (decrease) of light in the upper half of the column, which is observed in the Schlieren signals of figure \ref{fig:schlierenfrequencydependent} for different laser detuning frequencies.\\
For detuning frequencies $ \Delta \nu_L \ge \pm 9 \, \textnormal{GHz}$ the Schlieren signal starts to fade. Here the pump does not sufficiently excite the vapor (due to the larger difference between the laser and the transition frequency) and the deflection of the Schlieren beam is too small to be detected. Approaching the transition frequency the Schlieren signal becomes stronger, since the excitation by the pump beam increases and the difference in the refractive indices between non-excited and excited vapor also increases. With this setup the laser frequency cannot approach the transition frequency arbitrarily close: If the laser frequency is too close to the transition frequency the attenuation of the Schlieren beam by the Rb vapor becomes too strong and the Schlieren beam undetectable, as depicted in the central image of figure \ref{fig:schlierenfrequencydependent}, obtained with a detuning frequency $ \Delta \nu_L = +1 \, \textnormal{GHz}$.\\
\begin{figure}[htb!]
\centering
		\includegraphics[width =0.7\columnwidth,trim={2cm 6.8cm 2cm 6cm},clip]{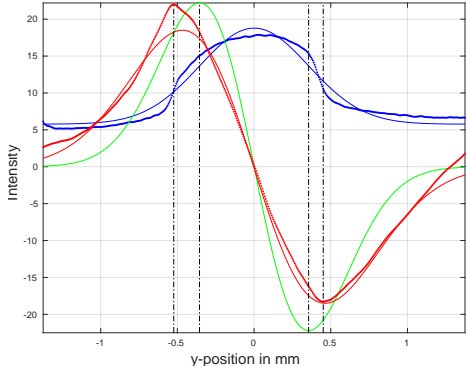}
		\caption{Lineouts integrated along the x-coordinate of the images as shown in \ref{fig:schlierenshadowgraphycomparison4conf_knifeedge}, recorded at high density ($T = 150 ^{\circ}C$) and a laser detuning frequency $ \Delta \nu_L = - 1.9 \, \textnormal{GHz}$: the lineout of the fluorescence image b, multiplied by a factor $15$ (blue dotted line), its Gaussian fit (blue solid line), the fit's derivative (green solid line) and of the Schlieren signal (red dotted line) with a Gaussian derivative fit (red solid line)}
\label{fig:radiuslineouts}
\end{figure}
From the four images (see figure \ref{fig:schlierenshadowgraphycomparison4conf_knifeedge})  the radius of the excitation column is estimated. Figure \ref{fig:radiuslineouts} shows lineouts integrated along the x-coordinate of the images: The blue dotted line of the fluorescence image (as e.g. in figure \ref{fig:schlierenshadowgraphycomparison4conf_knifeedge} b) is multiplied by a factor of $15$ for a better size comparison with the Schlieren signal, the red dotted line. We see again that the intensity of the Schlieren signal is significantly higher than the signal obtained by recording the fluorescence.\\
The laser beam has a Gaussian transverse profile. Assuming that the pump laser intensity is low enough not to reach saturation of the D2 transition, the depletion of the Rb ground state and thus the fluorescence signal is expected to follow the transverse pump beam profile. Since the Abel transform of a Gaussian is again a Gaussian the fluorescence light is fitted with a Gaussian function (blue solid curve). 
Using equation \ref{eq:refractiveindexvapor} and a Taylor expansion (since in the saturated case $N_j = N_i$ for a Rb density of $n = 1\cdot 10^{15} \textnormal{cm}^{-3}$ and a detuning frequency of $\Delta \nu_L=-1.9 \, \textnormal{GHz}$ the refractive index $n_{vapor} \approx 1.005$) the variation of the index of refraction of the Rb is $n_{vapor}(r) \approx 1 + \alpha /2 \cdot N_i(r) $ with a constant $\alpha$, thus also a Gaussian function. Schlieren imaging produces a signal that is proportional to the first spatial derivative of the refractive index $\partial n/\partial y$ of the transparent object for a horizontal knife edge \cite{schlierenSpringer}. Thus the Schlieren signal (red dotted points) is fitted with the derivative of a Gaussian function (red solid curve). For comparison, additionally the derivative of the Gaussian fit (blue solid line) is plotted (green curve).\\
The dashed black vertical lines indicate the extremes of the green curve and of the red data points of the Schlieren signal. The similar extreme positions of the two curves support the assumption that the radius of the excited Rb vapor column can be measured with Schlieren imaging. We determine the radius from the distance of the extremes. The results obtained with Schlieren imaging with a laser detuning frequency of $\Delta \nu_L = - 1.9 \, \textnormal{GHz}$ show an excitation column diameter of $d = 0.97 \, \textnormal{mm}$ on the CCD camera. The fluorescence image indicates a diameter of $d = 0.71 \, \textnormal{mm}$. Including the optical magnification of the two lenses of the Schlieren setup ($M = 4/3$) this corresponds to an excitation radius of $r = 0.36 \, \textnormal{mm}$ for the Schlieren signal and $r = 0.27 \, \textnormal{mm}$ for the fluorescence signal. Thus the excited column radius, determined with Schlieren imaging, is close to the one measured from the fluorescence light.\\
\section{Conclusions}
\label{sec:conclusion}
We demonstrate, using Schlieren imaging, that the radius of an excited column of Rb vapor can be determined with an accuracy of approximately $30 \%$. We used split versions of a laser beam tuned to near the D2 Rb line for the measurement. 
For a laser frequency very close to the D2 Rb transition line a large enough number of Rb atoms were excited and a Schlieren signal of the column was recorded. The radius of the excitation column for a laser detuning frequency of $\Delta \nu_L = - 1.9 \, \textnormal{GHz}$ was determined with this method as $r = 0.36 \, \textnormal{mm}$.\\
In the AWAKE experiment a short ($\tau \approx 100 \, \textnormal{fs}$), powerful ($P \approx 4 \textnormal{TW}$) laser pulse with a central wavelength of $\lambda \approx 780 \, \textnormal{nm}$ pumps and ionizes the Rb vapor. The laser bandwidth is broad enough (FWHM $10 \, \textnormal{nm}$) to also pump optical transitions from the upper level of the D2 line as well as the D1 line (see figure \ref{fig:rbenergylevels}). This will depopulate the population of the upper state of the D2 transition, which will modify the index of refraction at large pump intensities, possibly bringing it to near one, when the ground state population is fully depleted. This may make the transition from a depleted ground state population region ($n_{vapor} \simeq 1$) to an ionized region ($n_{plasma} =  \sqrt{1-\omega_{pe}^2/\omega_L^2}$) difficult to observe using Schlieren imaging, since $\omega_L^2 \gg \omega_{ij}^2$. Further experiments will explore this case.\\
We want to repeat the measurement tuning the imaging laser to the wavelength of the Rb transition line from the first to the second excited state ($\lambda \approx 776 \, \textnormal{nm}$, see figure \ref{fig:rbenergylevels}) in order to resolve transitions from a region of Rb in the first excited state, that is expected to surround the plasma column when the column is excited by the short laser pulse.\\
In contrast to the steady state process of the shown Schlieren experiment of a Rb excitation column created by a CW laser, the Rb excitation and ionization by the short pulsed laser is a dynamic process. We can potentially use a gated camera to time resolve the evolution of the excited/ionized Rb column.\\

\section{Acknowledgments}
\label{sec:Acknowledgments}
This work is sponsored by the Wolfgang Gentner Program of the German Federal Ministry of Education and Research.
\appendix

\end{document}